# FIVE SUPERNOVA SURVEY GALAXIES IN THE SOUTHERN HEMISPHERE. I. OPTICAL AND NEAR-INFRARED DATABASE

A. A. Hakobyan[1], A. R. Petrosian[1], G. A. Mamon[2], B. McLean[3], D. Kunth[2], M. Turatto[4], E. Cappellaro[5], F. Mannucci[6], R. J. Allen[3], N. Panagia[3,4,7], M. Della Valle[8]

The determination of the supernova (SN) rate is based not only on the number of detected events, but also on the properties of the parent galaxy population. This is the first paper of a series aimed at obtaining new, refined, SN rates from a set of five SN surveys, by making use of a joint analysis of near-infrared (NIR) data. We describe the properties of the 3838 galaxies that were monitored for SNe events, including newly determined morphologies and their DENIS and POSS-II/UKST *I*, 2MASS and DENIS *J* and $K_s$ and 2MASS *H* magnitudes. We have compared 2MASS, DENIS and POSS-II/UKST *IJK* magnitudes in order to find possible systematic photometric shifts in the measurements. The DENIS and POSS-II/UKST *I* band magnitudes show large discrepancies (mean absolute difference of 0.4 mag), mostly due to different spectral responses of the two instruments, with an important contribution (0.33 mag *rms*) from the large uncertainties in the photometric calibration of the POSS-II and UKST photographic plates. In the other wavebands, the limiting near infrared magnitude, morphology and inclination of the galaxies are the most influential factors which affect the determination of photometry of the galaxies. Nevertheless, no significant systematic differences have been found between of any pair of NIR magnitude measurements, except for a few percent of galaxies showing large discrepancies. This allows us to combine DENIS and 2MASS data for the *J* and $K_s$ filters.

## 1. Introduction

The supernova (SN) rate normalized to the stellar mass of the parent galaxies [1] contains unique information on the star formation history, stellar content and chemical evolution of the galaxies [2]. The accuracy of the empirical determination of the SN rate depends strongly not only on the quality

___________

[1] V.A.Ambartsumian Byurakan Astrophysical Observatory, Byurakan 0213, Armenia, e-mail: hakartur@rambler.ru

[2] Institut d'Astrophysique de Paris (UMR 7095: CNRS & UPMC), 98 bis Boulevard Arago, F-75014 Paris, France

[3] Space Telescope Science Institute, 3700 San Martin Drive, Baltimore, MD 21218, USA

[4] INAF, Osservatorio Astrofisico di Catania Via Santa Sofia 78, I-95123 Catania, Italy

[5] INAF, Osservatorio Astronomico di Padova, Vicolo dell'Osservatorio 5, I-35122 Padova, Italy

[6] CNR – IRA, Largo E. Fermi 5, I-50125 Firenze, Italy

[7] Supernova Ltd., OYV #131, Northsound Road, Virgin Gorda, British Virgin Islands

[8] INAF, Osservatorio Astronomico di Capodimonte, salita Moiariello 16, I-80131 Napoli, Italy

and duration of the surveys, providing the total number of SNe, but also on the level of knowledge on the parent galaxy population.

Most of the information of the SN rate in the local Universe is based upon five photographic and visual SN surveys (FSS), described in [3,4] and on the still ongoing LOTOSS project [5]. The quality of the resulting SN rates depends strongly on the quality of the information on accurate distances, morphologies, luminosities, colors and inclinations of the monitored galaxies [3,6]. To obtain meaningful information, it is necessary to relate the number of detected SNe to the "size" of the target galaxies, normalizing the SN rate to one chosen parameter of the parent galaxies. This choice plays a crucial role in the results and on its physical meaning. Historically, the parameter most commonly used is the luminosity in the optical $B$ band [7], assumed to be a measure of the galaxy stellar mass [8]. Currently it is widely believed that even if $B$ band luminosity would be an acceptable measure of the stellar mass in galaxies of a given morphological type, it is very poor tracer of stellar mass along the whole Hubble sequence [1]. The near-infrared (NIR) wavebands (from 0.8 to 2.2 $\mu m$) are much better tracers of the stellar mass of galaxies [9,10], hence the normalization of the SN rate to NIR luminosities is more valuable [11]. The recent release of the Two Micron All Sky Survey (2MASS, [12]), including catalogues of NIR magnitudes for hundred thousand of galaxies, makes such an approach possible [1]. Another very large list of NIR magnitudes is provided by the Deep Near-Infrared Southern Sky Survey (DENIS), which was first presented by Epchtein et al. [13,14]. Although the final DENIS galaxy catalogue is still in preparation [15], NIR apparent magnitudes as well as diameters have been measured from the DENIS images by Paturel et al. ([16]; hereafter P05), for several hundred thousand galaxies in the Lyon Extragalactic Database (LEDA) catalog, which is part of the HyperLEDA database.

In the series of papers currently in preparation we aim to study the properties of galaxies targeted by the FSS. In particular, we are interested in computing the determination of the rates of various types of SNe normalized to the stellar mass of the parent galaxies as inferred from both the 2MASS and DENIS NIR photometry. To properly carry out these studies, we have created the database of FSS galaxies identified in the field of DENIS survey. This database for survey galaxies includes newly determined morphologies and newly measured apparent blue and red magnitudes, angular diameters, axial ratios and position angles, 2MASS $J$, $H$ and $K_s$ and DENIS $I$, $J$ and $K_s$ magnitudes, as well as activity classes from the NASA/IPAC Extragalactic Database (NED) and literature and numbers of neighboring objects in a circle of radius 50 kpc. The database includes also extensive notes summarizing a large set of information about different properties of galaxies, allowing the measurement of the rate as a function, for example, of radio flux [17] and environment [18]. The creation of this homogeneous database is aimed to support all future studies as well to minimize possible selection effects and errors which often arise when information for studied objects is selected from different sources and catalogues.

In this first paper we present results of a comparative study of our database, paying special attention to DENIS data. In Section 2 we present input catalogue and determined optical parameters; in Section 3 DENIS and related data for the sample galaxies will be presented. In Section 4 a comparison of the DENIS magnitudes with the 2MASS, P05 and photographic data will be discussed. Section 5 will summarize the main conclusions of this study. In this paper we have assumed a value for the Hubble constant of $H_0 = 75$ km s$^{-1}$ Mpc$^{-1}$.

## 2. The Five SN surveys (FSS) and their optical parameters

In our studies we will use the SN and galaxy lists from [4], which was created by joining the logs of FSS [3], namely the Asiago [19], Crimea [20], Cálan-Tololo [21] and OCA photographic surveys [22] and the visual search by Evans [23]. The FSS covers both Northern and Southern hemispheres and monitored over 10000 galaxies listed in the Third Reference Catalogue of Bright Galaxies ([24]; hereafter RC3). The measurement of the SN rate requires the knowledge of the recession velocity, the morphological type, the luminosity and the axial ratio for spirals. The total number of galaxies with all these parameters in the five surveys is 7773, containing 136 SNe events [3,4]. In the area covered by the DENIS survey (see Section 3), the number of the FSS galaxies is 3838 and the number of detected SN events is 74 in 53 galaxies. This is the sample studied in this series of papers.

The object morphologies and axial ratios were determined from blue and red images on photographic plates: the Palomar Schmidt plates (POSS-II) for the northern hemisphere and the UK Schmidt (UKST) plates for the southern hemisphere. Approximately 93% of the 3838 galaxies are located on UKST survey plates and about 7% on POSS-II plates. All plates used in our analysis have been digitized at STScI using the modified PDS microdensitometer with a pixel size of 15 $\mu m$ (1$''$0). The optical parameters (morphologies and axial ratios) for the FSS galaxies were then extracted as follows: 10' x 10' regions centered on each FSS galaxy were extracted from the POSS-II IIIa-J ($\lambda_{eff}$ ~ 4800Å) and IIIa-F ($\lambda_{eff}$ ~ 6500Å) plates and from the UKST IIIa-J ($\lambda_{eff}$ ~ 4800Å) and IIIa-F ($\lambda_{eff}$ ~ 6300Å) plates. Using these images, we determined the morphologies of the sample galaxies (A.A.H.) and measured blue and red apparent magnitudes (B.M.), major and minor diameters (A.A.H.), position angles (A.A.H.) and calculated the apparent number of neighboring galaxies within a 50 kpc radius (A.A.H.), based on the known published redshifts mostly from NED.

The method to determine the morphological classification of our galaxies and its accuracy is discussed in detail elsewhere [25,26]. According to our morphological determinations, 16% of the sample galaxies are elliptical, 24% are S0 type objects, 14% are Sa/Sab, 23% are Sb/Sbc, 21% are Sc/Scd and remaining 2% are Sd/Sm type spirals and irregular galaxies. The blue and red apparent magnitudes of the galaxies were homogeneously measured on these plates at the approximately 25.3 mag arcsec$^{-2}$ isophote in both wavebands, corresponding to roughly 3 times the background *rms* noise. The method and accuracy of our blue and red apparent magnitudes measurements are discussed in detail in [25,26]. Blue apparent magnitudes of sample galaxies cover the range from 2.7 mag (Small Magellanic Cloud) to 17.2 mag with an average value of 14.0 ± 0.8 mag. The major and minor angular diameters of the FSS galaxies were also measured within the same isophote (see [25,26]) and were used to obtain the axial ratios. The FSS galaxies have a homogeneous distribution of axial ratios, with an average value of 0.58 and a standard deviation of 0.23. The position angles (PA) of the major axes of the galaxies were determined for the same mean limiting surface brightness. Counts of neighboring galaxies were done for all FSS galaxies with known redshifts by projecting a circle of 50 kpc radius on a 10' x 10' digitized field of each galaxy. All galaxies detected within this circle were counted if their angular sizes differed from that of the sample galaxy by no more than factor of 2 [25-27]. Because of several technical difficulties and uncertainties (extremely large angular field size, etc.) the counts of neighboring galaxies were only determined for galaxies with redshifts greater than 0.005.

The selection of active or star-forming (A/SF) galaxies among sample objects have been made by cross-checking this sample with the all known possible sources of active or star-forming galaxies. The cross-checking analysis includes well known optical surveys of A/SF galaxies, e.g., Markarian [25,28], Kiso [29] and others, as well as radio (e.g., [30]) surveys, which cover the DENIS survey area. Cross-checking also includes the lists of known peculiar morphological structure galaxies (e.g., PGC 438; [31]). The inclusion of galaxies with peculiar morphologies, which very often are the outcomes of merging or close interaction, follows the common belief that, in general, star formation or nuclear activity is enhanced in these galaxies [32]. In this respect, the A/SF galaxies sample includes also objects that, according to our study, have been morphologically classified as mergers or close interacting systems. Normal galaxies are those that are not included in any list of A/SF galaxies or are not X-Ray or radio sources and have no recorded peculiar morphological and other properties. Spectral data for about 8% of the sample galaxies have been presented in the 5$^{th}$ data release of the Sloan Digital Sky Survey (SDSS DR5). This information also was used in the classification of an object as A/SF or normal. In total, among 3838 sample galaxies, we find that 688 are A/SF galaxies, i.e. about 18% of the entire sample.

## 3. NIR data for sample galaxies

The 2MASS survey was carried out in both Northern and in Southern hemispheres in the $J$ (1.25 $\mu m$), $H$ (1.65 $\mu m$) and $K_s$ (2.12 $\mu m$) bands [33]. The final catalogue of 2MASS is available online (http://www.ipac.caltech.edu/2mass/). The cross-identification is accepted when there is only one candidate within a distance of $r = 10"$ from each galaxy and when the agreement between coordinates is better than 5". It includes only 2MASS detections with confusion ([jhk]_flg) flags 0, 1 and 2 and contamination/confusion (cc_flg) flag of 0 or Z and reveals 2955 identifications at $J$ band, 2935 identifications at $H$ band, and 2968 identifications at $K_s$ band. From 2MASS Kron elliptical aperture magnitudes were also extracted.

The DENIS survey is similar to 2MASS, with one million images in the $I$ (Gunn-$i$) band – at 0.8 $\mu m$, as well as the $J$ and $K_s$ bands. DENIS covers a sky area between declination +2° in the North and −88° in the South (http://cdsweb.u-strasbg.fr/denis.html). Each elementary image is 12' x 12' with a pixel size of 1". The integration time is 9 seconds. The sequence of observations is made at the given right ascension for a wide range of declination (30°). This arrangement is called the strip. It contains 180 elementary images with 1' overlap on each side.

The final catalogue of DENIS galaxies to be generated from a blind extraction of DENIS images is in preparation [15]. A preliminary extraction of DENIS galaxies was performed by Paturel et al. (P05) for the 750 thousand galaxies already in LEDA. Moreover, one of us (G.A.M.) extracted DENIS $I$, $J$ and $K_s$ Kron magnitudes [34] for the 3838 sample galaxies, using the SExtractor software [35]. In this second extraction (hereafter "Ours"), DENIS detections with SExtractor object flags between 0 and 3 [15] were considered. Cross-identification of the FSS galaxies with our DENIS measurements was conducted in the same way as for with 2MASS (see above). With this procedure, we extracted 2789, 2807 and 2595 galaxies in the DENIS $I$, $J$ and $K_s$ wavebands, respectively. Cross-identification of the FSS galaxies with the DENIS P05 catalogue reveals 2665, 2386 and 1654 identifications in the DENIS $I$, $J$ and $K_s$ bands, respectively. Object identification and data extraction from the P05 catalogue were done similarly as for 2MASS and (our extraction of) DENIS.

Since the photographic POSS-II and UKST surveys were also carried out in the $I$ band (IV-N emulsion), using the RG9 filter ($\lambda_{eff}$ ~ 8500Å) for POSS-II and the RG715 filter ($\lambda_{eff}$ ~ 7900Å) for UKST, we also used these data for comparison with DENIS $I$ band observations. One of us (B.M.) repeated the same analysis for the FSS galaxies on the IV-N plates as he performed on the IIIa-J and IIIa-F plates. Cross-identification of the FSS galaxies with the POSS-II and UKST $I$ band images reveals 3792 identifications. The $I$ band magnitudes were measured with a similar method as for POSS-II and UKST $J_{pg}$ and $F_{pg}$ magnitudes. Again, the cross-identification of the FSS galaxies with the POSS-II and UKST catalogues (hereafter POSS/UKST) was done in the same fashion as with 2MASS and DENIS.

The number of galaxies present in both DENIS and 2MASS catalogs are 1818 in the $J$ band and 981 in the $K_s$ band. For the DENIS $I$ band, we compared our measurements with P05 data as well with POSS/UKST photographic $I$ band magnitudes. The intersection of the two DENIS $I$ band samples (Ours and P05) contains 2207 entries. The intersection of Ours and POSS/UKST $I$ band magnitudes comparison sample includes 2771 galaxies. For all of these galaxies, a homogeneous set of optical parameters exists.

## 4. Results

Using samples heavily dominated by stars, Carpenter [36] and Cabrera-Lavers & Garzón [37] concluded that there are no significant (greater than 0.01 mag) differences between the 2MASS and DENIS photometry in the $J$ and $K_s$ filters. Nevertheless, both studies found a small fraction of stars with greater than 0.5 mag differences between DENIS and 2MASS (5% from 200 stars [36] and less than 3% from 26253 stars [37]). For diffuse objects, P05 (see also [38]) found that their Kron magnitude differs from the total 2MASS magnitude by $J$(P05) − $J$(2MASS) = − 0.023 ± 0.210 ($N$ = 48439) and $K_s$(P05) − $K_s$(2MASS) = 0.069 ± 0.350 ($N$ = 21943), hence the photometry from both surveys is in good agreement, although magnitude differences larger than 1 mag can be found, in both bands, for a large number of galaxies. P05 stressed that the contamination by superimposed objects (stars or companion galaxies) can be the major cause of such differences. Below, separately for $I$, $J$ and $K_s$ bands the results of comparative studies of the near-IR photometry and related optical parameters are presented.

**4.1. The $I$-Band.** In this band, our DENIS magnitudes are compared with P05 DENIS measurements and POSS-II and UKST results. The percentage distributions of the absolute differences between our and P05 magnitudes and our DENIS and POSS/UKST measurements are shown in Fig. 1, with statistics presented in Tables 1 and 2. The dependence between our and P05 DENIS magnitude differences can be fit in with a linear regression, which has following form (see Table 1):

$$I(\text{P05}) - I(\text{Ours}) = - (0.250 \pm 0.066) + (0.015 \pm 0.005)\, I(\text{Ours}). \qquad (1)$$

The mean difference between the P05 and our DENIS measurements is 0.07 ± 0.26, while the mean absolute difference is 0.16 ± 0.22. For more than 55% of galaxies our and P05 magnitude differences are less than 0.1 mag and for more than 95% of the galaxies magnitude differences are less than 0.5 mag. But the magnitude difference is equal or larger than 1 mag for more than 1% of the galaxies. We paid special attention to all this cases, analyzing possible causes of the errors. In several cases, our DENIS measurements were questioned and which after comparison were corrected and in same cases, the P05 measurements seem questionable. Mostly, these cases are for galaxies in close interaction or for galaxies with one or more projected stars.

The mean difference between the POSS/UKST measurements is $0.13 \pm 0.49$, while the mean absolute difference is as large as $0.41 \pm 0.31$. For only about 16% of galaxies our DENIS and Schmidt plates *I* band magnitude differences are less than 0.1 mag and for no more than 67% of the galaxies magnitude differences are less than 0.5 mag. For more than 8% of the objects magnitude differences are equal or larger than 1 mag.

For only about 14% of galaxies P05 and Schmidt plates *I* band magnitude differences are less than 0.1 mag and for no more than 66% of the galaxies magnitude differences are less than 0.5 mag. For about 7% of the objects magnitude differences are equal or larger than 1 mag. When we include a color term, the fit becomes (see Table 2):

$$I(\text{POSS/UKST}) - I(\text{Ours}) = (1.978 \pm 0.113) - (0.127 \pm 0.008)\, I(\text{Ours}) - (0.653 \pm 0.022)\, (I-J)(\text{Ours}). \qquad (2)$$

The mean absolute difference is now reduced to $0.22 \pm 0.18$. Hence, most of the residuals, $(0.41^2 - 0.22^2)^{1/2} = 0.35$ mag, from the fit without color term between DENIS and POSS/UKST *I* band magnitudes are the result of the different spectral response curves of Gunn-*i* filter (DENIS) and IV-N emulsion plus RG9 (POSS-II) or RG715 (UKST) filters combinations. The remaining fairly high residuals ought to be caused by photometric calibration errors, which tend to be large for photographic plates and by the uncertainties and variations of the plate response curve can, which introduce non-negligible errors. Are the mean absolute residuals of 0.22 mag, which for Gaussian errors correspond to *rms* residuals of 0.33 mag, caused by the DENIS extractions or the POSS/UKST measurements? The analysis of overlaps of adjacent or repeated DENIS strips indicates that DENIS zero points are correct to 0.05 mag *rms*. On the other hand, P05 found that DENIS-P05 *I* matched deep measurements by Mathewson et al. [39] to within only 0.15 mag *rms*. This suggests that the *rms* errors on the individual POSS/UKST *I* band photometry are as large as $(0.33^2 - 0.15^2)^{1/2} = 0.29$ mag.

Table 2 also shows that there is almost no color term between the two sets of DENIS measurements (P05 and Ours), as expected, and the mean absolute residuals are small. However, surprisingly, Table 2 indicates a very small color term between POSS/UKST and DENIS (P05), in contrast with the very strong color term between POSS/UKST and DENIS (Ours). The fractions of bad matches (absolute difference greater than 0.5 mag) with color term are less than 6% for all pairs in *I* band. We therefore adopt the DENIS (Ours) measurements, except for the galaxies with no DENIS measurements, for which we adopt the POSS/UKST *I* band data, after transformation to the DENIS *I* system, using eq. (2).

TABLE 1. Photometric Comparisons without Color Terms

| $A_1$ | $A_2$ | $a_1$ | $b_1$ | $N$ | $\langle A_1 - A_2 \rangle$ | $\langle |A_1 - A_2| \rangle$ |
|---|---|---|---|---|---|---|
| (1) | (2) | (3) | (4) | (5) | (6) | (7) |
| $I$(POSS/UKST) | $I$(Ours) | $0.136 \pm 0.105$ | $-0.022 \pm 0.008$ | 2768 | $0.13 \pm 0.49$ | $0.41 \pm 0.31$ |
| $I$(P05) | $I$(Ours) | $-0.250 \pm 0.066$ | $0.015 \pm 0.005$ | 2207 | $0.07 \pm 0.26$ | $0.16 \pm 0.22$ |
| $I$(POSS/UKST) | $I$(P05) | $1.075 \pm 0.135$ | $-0.091 \pm 0.011$ | 2207 | $-0.05 \pm 0.57$ | $0.43 \pm 0.37$ |
| $J$(2MASS) | $J$(Ours) | $-0.509 \pm 0.057$ | $0.050 \pm 0.005$ | 1818 | $-0.06 \pm 0.22$ | $0.14 \pm 0.18$ |
| $J$(P05) | $J$(Ours) | $-0.025 \pm 0.073$ | $-0.010 \pm 0.006$ | 1818 | $0.14 \pm 0.27$ | $0.20 \pm 0.23$ |
| $J$(2MASS) | $J$(P05) | $0.292 \pm 0.085$ | $-0.008 \pm 0.008$ | 1818 | $-0.20 \pm 0.33$ | $0.26 \pm 0.28$ |
| $K$(2MASS) | $K$(Ours) | $0.662 \pm 0.111$ | $-0.080 \pm 0.011$ | 981 | $0.17 \pm 0.33$ | $0.25 \pm 0.27$ |
| $K$(P05) | $K$(Ours) | $0.919 \pm 0.165$ | $-0.105 \pm 0.016$ | 981 | $0.17 \pm 0.49$ | $0.32 \pm 0.41$ |
| $K$(2MASS) | $K$(P05) | $1.274 \pm 0.126$ | $-0.124 \pm 0.012$ | 981 | $-0.01 \pm 0.40$ | $0.24 \pm 0.32$ |

$$A_1 - A_2 = a_1 + b_1 A_2$$

(1) – Waveband (survey)
(2) – Reference waveband (survey)
(3), (4) – Estimated parameters and standard errors
(5) – Number of pairs
(6) – Mean difference and standard deviation from the database
(7) – Mean absolute difference and standard deviation from the database

TABLE 2. Photometric Comparisons with Color Terms

| $A_1$ | $A_2$ | $B$ | $a_2$ | $b_2$ | $c_2$ | $N$ | $\langle A_1 - A_2 \rangle$ | $\langle |A_1 - A_2| \rangle$ |
|---|---|---|---|---|---|---|---|---|
| (1) | (2) | (3) | (4) | (5) | (6) | (7) | (8) | (9) |
| $I$(POSS/UKST) | $I$(Ours) | $I$(Ours) $-$ $J$(Ours) | $1.978 \pm 0.113$ | $-0.127 \pm 0.008$ | $-0.653 \pm 0.022$ | 2567 | $-0.14 \pm 0.25$ | $0.22 \pm 0.18$ |
| $I$(P05) | $I$(Ours) | $I$(Ours) $-$ $J$(Ours) | $-0.414 \pm 0.080$ | $0.026 \pm 0.006$ | $0.029 \pm 0.016$ | 2077 | $-0.07 \pm 0.02$ | $0.07 \pm 0.02$ |
| $I$(POSS/UKST) | $I$(P05) | $I$(P05) $-$ $J$(P05) | $1.475 \pm 0.209$ | $-0.129 \pm 0.018$ | $-0.021 \pm 0.012$ | 1642 | $-0.13 \pm 0.15$ | $0.16 \pm 0.11$ |
| $J$(2MASS) | $J$(Ours) | $J$(2MASS) $-$ $K$(2MASS) | $-0.077 \pm 0.094$ | $0.008 \pm 0.009$ | $0.035 \pm 0.006$ | 864 | $0.05 \pm 0.05$ | $0.06 \pm 0.04$ |
| $J$(P05) | $J$(Ours) | $J$(2MASS) $-$ $K$(2MASS) | $-0.134 \pm 0.154$ | $-0.000 \pm 0.014$ | $-0.004 \pm 0.010$ | 864 | $-0.14 \pm 0.01$ | $0.14 \pm 0.01$ |
| $J$(2MASS) | $J$(P05) | $J$(2MASS) $-$ $K$(2MASS) | $1.488 \pm 0.144$ | $-0.127 \pm 0.013$ | $0.108 \pm 0.010$ | 864 | $0.19 \pm 0.11$ | $0.19 \pm 0.10$ |
| $K$(2MASS) | $K$(Ours) | $J$(2MASS) $-$ $K$(2MASS) | $1.596 \pm 0.140$ | $-0.158 \pm 0.013$ | $-0.094 \pm 0.009$ | 864 | $-0.15 \pm 0.12$ | $0.16 \pm 0.11$ |
| $K$(P05) | $K$(Ours) | $J$(2MASS) $-$ $K$(2MASS) | $1.484 \pm 0.223$ | $-0.150 \pm 0.020$ | $-0.068 \pm 0.014$ | 864 | $-0.15 \pm 0.11$ | $0.16 \pm 0.10$ |
| $K$(2MASS) | $K$(P05) | $J$(2MASS) $-$ $K$(2MASS) | $2.489 \pm 0.161$ | $-0.230 \pm 0.015$ | $-0.121 \pm 0.010$ | 864 | $-0.00 \pm 0.18$ | $0.14 \pm 0.11$ |

$$A_1 - A_2 = a_2 + b_2 A_2 + c_2 B$$

(1) – Waveband (survey)
(2) – Reference waveband (survey)
(3) – Reference color
(4), (5), (6) – Estimated parameters and standard errors
(7) – Number of pairs
(8) – Corrected mean difference and standard deviation
(9) – Corrected mean absolute difference and standard deviation

Whereas erroneous photometry can be caused by misidentification of a galaxy or by contamination by stars or nearby galaxies, the photometric differences might also depend on the global galaxy parameters, particularly their morphology, apparent magnitude, angular size and inclination. To study this possible relationship we applied Multivariate Factor Analysis (MFA, [40]). We choose the following initial variables: the two magnitude differences, our magnitude, morphology, angular diameter and axial ratio.

In order to present each initial variable with the smallest number of common factors we apply to the two factors the orthogonal rotation that is variance maximizing (Varimax [41]), which essentially turns the MFA into a Principal Component Analysis (PCA). Table 3 shows the factor loadings, i.e. the correlation coefficients between the initial variables and the factors. The variance accumulated by the first two factors is 88%. Adopting $r \sim 0.5$ as correlation threshold, the first factor, which accounts for 70% of the common variance, is the combination of the $I$ band magnitude, morphology, axial ratio of the galaxies and both $I$ band magnitude absolute differences, with fainter $I$ band magnitude, later morphological type and less inclined galaxies show highest absolute differences in photometry. The correlation between $I$ band magnitudes and differences in photometry obviously can be caused by the correlation between limiting magnitudes and internal standard errors for both P05 and our DENIS photometry [15]. Photometric detection of the low surface brightness outer structures of late type galaxies, especially when seen face on, is problematic as it particularly affects the determination of the Kron radius within which the Kron magnitude is measured [34]. For such objects, different approaches to the technique of the photometry adopted by P05 and [15] can lead to large magnitude differences.

TABLE 3. Correlation Coefficients Between Initial Variables and Varimax-Rotated MFA Factors for $I$ Band Data

|  | Factors | |
| --- | --- | --- |
|  | 1 | 2 |
| $I$(DENIS – Ours) | 0.930 | –0.302 |
| morphology | 0.693 | 0.072 |
| angular diameter | –0.009 | 0.997 |
| axial ratio | 0.975 | –0.009 |
| $|I$(Ours) – $I$(P05)$|$ | 0.982 | 0.022 |
| $|I$(Ours) – $I$(POSS/UKST)$|$ | 0.980 | –0.006 |
| cumulative variance in % | 70.3 | 88.5 |

Factor 2, which accounts for 18.3% of the common variance, correlates only with the angular diameter of the galaxies. Since according to MFA method the factors are independent, we conclude that angular diameter does not contribute significantly to erroneous photometry.

**4.2. The *J*-band.** In this band our DENIS magnitudes are compared with the P05 DENIS and 2MASS measurements for 1818 galaxies. The distributions of absolute magnitude differences are shown in Fig. 2 with statistics given in Table 1 and 2. The agreement between our and P05 DENIS $J$ band magnitude is very good. The mean magnitude difference is $0.14 \pm 0.27$ and the mean absolute value of the magnitude difference is $0.20 \pm 0.23$. For only 37% of galaxies our and P05 $J$ band magnitude differences are less than 0.1 mag, and for more than 93% of the galaxies magnitude differences are equal or less than 0.5 mag. For more than 2% of the objects magnitude differences are equal or larger than 1 mag.

Our DENIS measurements are better matched to the 2MASS ones. The mean magnitude difference is $-0.06 \pm 0.22$ and the mean absolute difference is $0.14 \pm 0.18$. For 59% of galaxies our and 2MASS $J$ band magnitude differences are less than 0.1 mag and for more than 96% of the galaxies magnitude differences are equal or less than 0.5 mag. For less than 1% of the objects magnitude differences are equal or larger than 1 mag.

The match between the DENIS-P05 and 2MASS $J$ band photometries is worse. The mean magnitude difference is $-0.20 \pm 0.33$ and the mean absolute difference is $0.26 \pm 0.28$. For only 24% of galaxies P05 and 2MASS $J$ band magnitude differences are less than 0.1 mag and for the 89% of the galaxies magnitude differences are equal or less than 0.5 mag. For the 3% of the objects magnitude differences are equal or larger than 1 mag. It is surprising that among the two DENIS sets of photometric measurements and the 2MASS photometry, the closest match is between our DENIS photometry and the 2MASS photometry, rather than between both sets of DENIS measurements (both without and with the inclusion of a color term in the photometric equation).

We repeat the MFA analysis in the same way as was done in the $I$ band. Table 4 shows the factor loadings. Accumulated variance by the first two MFA factors is 90%. The first factor, which accounts for 75% of the common variance, is the combination of the $J$ band magnitude, morphology, axial ratio of the galaxies and all three $J$ band magnitude absolute differences, with fainter apparent $J$ band magnitude, later morphological type and less inclined galaxies show highest absolute differences in photometry. These trends are thus similar to those in the $I$ band and can be interpreted in the same way. The second factor, which accounts for 15% of the common variance, correlates only with the angular diameter of the galaxies; the angular diameter is not a significant factor which contributes on erroneous photometry.

TABLE 4. Correlation Coefficients Between Initial Variables and Varimax-Rotated MFA Factors for $J$ Band Data

|  | Factors | |
|---|---|---|
|  | 1 | 2 |
| $J$(DENIS – Ours) | 0.937 | –0.270 |
| morphology | 0.691 | 0.131 |
| angular diameter | –0.003 | 0.995 |
| axial ratio | 0.979 | –0.024 |
| $|J$(Ours) – $J$(P05)$|$ | 0.988 | 0.012 |
| $|J$(Ours) – $J$(2MASS)$|$ | 0.989 | –0.006 |
| $|J$(P05) – $J$(2MASS)$|$ | 0.990 | 0.000 |
| cumulative variance in % | 75.0 | 90.4 |

**4.3. The $K_s$-Band.** Our DENIS magnitudes are compared with the P05 DENIS and 2MASS measurements, just as for the $J$ band. In the $K_s$ band, there are 981 galaxies in common, with reliable flags, between our DENIS sample, the P05 DENIS sample and the 2MASS sample. The distributions of the absolute differences between our and P05 magnitudes, our and 2MASS and P05 and 2MASS measurements are presented respectively in Fig. 3. The match between the two sets of DENIS $K_s$ magnitudes is much worse than for the $J$ band. The mean difference between our and P05 magnitudes is $0.17 \pm 0.49$, while the mean absolute difference is $0.32 \pm 0.41$. For only 31% of the galaxies do our and P05 $K_s$ band magnitudes agree to better than 0.1 mag, but for over 83% of the galaxies, the magnitude differences are equal or less than 0.5 mag. For 7% of the objects, the magnitude differences are greater than 1 mag.

Our DENIS photometry is not better matched to the 2MASS magnitudes. The mean difference between our DENIS and the 2MASS measurements is $0.17 \pm 0.33$, while the mean absolute difference is $0.25 \pm 0.27$. 34% of the galaxies in common between our DENIS and 2MASS, the $K_s$ band magnitude differences are less than 0.1 mag, while for 88% of the galaxies the magnitude differences are less than 0.5 mag. For 4% of the objects, the magnitude differences are greater than 1 mag.

Similarly, the mean difference between the DENIS-P05 and 2MASS measurements is $-0.01 \pm 0.40$ and mean absolute difference is $0.24 \pm 0.32$. 39% of galaxies have P05 and 2MASS $K_s$ band magnitude differences of less than 0.1 mag, and for the 89% of the galaxies the magnitude differences are less than 0.5 mag. For 4% of the galaxies, the magnitude differences are greater than 1 mag. The much larger photometric discrepancies between DENIS and 2MASS in the $K_s$ band relative to that in the $J$ band reflects the strong $K_s$ background of DENIS, which is almost entirely caused by thermal emission from the instrument [42]. Based on the same DENIS data, the P05 measurements also show large differences with ours. This is again due to the uncertainties in the subtraction of the very high $K_s$ background in the DENIS images. The fractions of bad matches (absolute difference greater than 0.5 mag) with color term are less than 1% for all pairs in both $J$ and $K_s$ bands.

As for the $I$ and $J$ bands, the $K_s$ band photometry can depend on the global parameters of the galaxies. To study this possible relationship we applied again the MFA method, with the same initial variables as for the $I$ and $J$ bands. Table 5 shows the factor loadings. The accumulated variance in the first two MFA factors is only 56.2%. The first MFA factor, which accounts only for 34.2% of the common variance, is now the combination of the three $K_s$ band magnitude absolute differences and (marginally) with the apparent magnitude and morphology, but not with the axis ratio as was the case in the $I$ and $J$ bands. We interpret this trend in the same way as for the $I$ and $J$ bands. However, the influence of apparent magnitude on the magnitude differences is less significant than for the $I$ and $J$ bands, probably because the differences are dominated by the large uncertainties in the background around these galaxies in the DENIS images. Also, the lower cumulative variance is compensated by a third factor involving the axial ratio.

TABLE 5. Correlation Coefficients Between Initial Variables and Varimax-Rotated MFA Factors for $K_s$ Band Data

|  | Factors | |
|---|---|---|
|  | 1 | 2 |
| $K_s$(DENIS) | 0.536 | −0.739 |
| morphology | 0.542 | −0.068 |
| angular diameter | 0.067 | 0.943 |
| axial ratio | −0.170 | −0.275 |
| $\|K_s(\text{Ours}) - K_s(\text{P05})\|$ | 0.881 | 0.140 |
| $\|K_s(\text{Ours}) - K_s(\text{2MASS})\|$ | 0.731 | 0.050 |
| $\|K_s(\text{P05}) - K_s(\text{2MASS})\|$ | 0.682 | 0.074 |
| cumulative variance in % | 34.2 | 56.2 |

The second factor, which accounts for 22% of the cumulative variance, correlates with the $K_s$ band magnitude and angular diameter of the galaxies, as brighter galaxies have larger angular diameters, which is obvious.

## 5. Conclusions

In this article, we report the creation of a database for the 3838 galaxies targeted in the FSS with photometry in 2MASS (*JHK*), DENIS (*IJK*) and the POSS II or UKST plates (*I*). The photometric measurements between these surveys are compared in pairs.

In the *I* band, there is excellent agreement between the P05 and our measurements of galaxies in the DENIS images, although for about 5% of the objects the P05 photometry differs by over 0.5 mag from ours. On the other hand, the agreement between the DENIS and POSS/UKST extractions is much worse, as one-third of the galaxies have photometry that disagree by over half a magnitude. This larger difference is mainly caused by the different instrumental spectral response curves used during DENIS and UKST observations. However, the high remaining residuals (0.33 mag *rms*) are principally caused by the uncertainties in the photometric calibration of the POSS-II and UKST photographic plates.

In the *J* band, the 2MASS photometry matches better our DENIS photometry than the P05 measurements of DENIS images. Only 4% of galaxies do our and 2MASS measurements differ by over 0.5 mag, while such discrepancies between DENIS-P05 and 2MASS photometry is found for 11% of the galaxies.

On the other hand, both magnitudes are in much worse agreement with the 2MASS ones, with 13% of galaxies differing by over 0.5 mag, which is most probably caused by the very high background of the DENIS $K_s$ band images.

Using a Multiple Factor Analysis, we find that the photometric errors are largest for faint, face-on late-type spirals, i.e. for low surface brightness galaxies, as one would expect. These factors are more important for the *I* and *J* bands than for the $K_s$ band, probably because of the very high background in the DENIS $K_s$ band images. In general, we have found that there is no significant difference between DENIS and 2MASS photometry and both data will be combined for the study of FSS galaxies.

# REFERENCES

1. F. Mannucci, M. Della Valle, N. Panagia et al., *Astron. Astrophys.,* **433**, 807, 2005.
2. F. Matteucci, N. Panagia, A. Pipino et al., *Mon. Notic. Roy. Astron. Soc.,* **372**, 265, 2006.
3. E. Cappellaro, M. Turatto, D. Yu. Tsvetkov et al., *Astron. Astrophys.,* **322**, 431, 1997.
4. E. Cappellaro, R. Evans, and M. Turatto, *Astron. Astrophys.,* **351**, 459, 1999.
5. A. V. Filippenko, W. D. Li, R. R. Treffers et al., in *"Small Telescope Astronomy in Global Scales"*, ASP Conference Series, Vol. 246, IAU Colloquium 183, 121, 2001.
6. S. van den Bergh, W. D. Li, and A. V. Filippenko, *Publ. Astron. Soc. Pacif.,* **117**, 773, 2005.
7. S. van den Bergh and G. A. Tammann, *Ann. Rev. Astron. Astrophys.,* **29**, 363, 1991.
8. G. A. Tammann, in *"Supernovae and Supernova Remnants",* Proceedings of the International Conference on Supernovae, Astrophysics and Space Science Library, Vol. 45, p.155, 1974.
9. E. F. Bell and R. S. de Jong, *Astrophys. J.,* **550**, 212, 2001.
10. J. J. Salzer, J. C. Lee, J. Melbourne et al., *Astrophys. J.,* **624**, 661, 2005.
11. M. Della Valle and M. Livio, *Astrophys. J. Lett.,* **423**, L31, 1994.
12. T. H. Jarrett, T. Chester, R. Cutri et al., *Astron. J.,* **125**, 525, 2003.
13. N. Epchtein, B. de Batz, L. Capoani et al., *The Messenger,* **87**, 27, 1997.
14. N. Epchtein, E. Deul, S. Derriere et al., *Astron. Astrophys.,* **349**, 236, 1999.
15. G. A. Mamon et al., *Astron. Astrophys.*, 2008, (in preparation).
16. G. Paturel, I. Vauglin, C. Petit et al., *Astron. Astrophys.,* **430**, 751, 2005, (P05).
17. M. Della Valle, N. Panagia, P. Padovani et al., *Astrophys. J.,* **629**, 750, 2005.
18. F. Mannucci, D. Maoz, K. Sharon et al., *Mon. Notic. Roy. Astron. Soc.,* **383**, 1121, 2008.
19. E. Cappellaro, M. Turatto, S. Benetti et al., *Astron. Astrophys.,* **268**, 472, 1993.
20. D. Yu. Tsvetkov, *Soviet Astr.,* **27**, 22, 1983.
21. M. Hamuy, J. Maza, M. M. Phillips et al., *Astron. J.,* **106**, 2392, 1993.
22. C. Pollas, in *"Supernovae"*, Proceedings of the 54[th] École d'été de physique théorique, session LIV, eds. S. A. Bludman, R. Mochkovitch, and J. Zinn-Justin, The Netherlands and North-Holland, Amsterdam, New York, p.769, 1994.
23. R. Evans, S. van den Bergh, and R. D. McClure, *Astrophys. J.,* **345**, 752, 1989.
24. G. de Vaucouleurs, A. de Vaucouleurs, H. G. Corwin et al., *"Third Reference Catalogue of Bright Galaxies",* Springer-Verlag, New York, 1991, (RC3).
25. A. Petrosian, B. McLean, R. J. Allen et al., *Astrophys. J. Suppl. Ser.,* **170**, 33, 2007.
26. A. Petrosian, B. McLean, R. Allen et al., *Astrophys. J. Suppl. Ser.,* **175**, 86, 2008.
27. I. D. Karachentsev, *Soobshch. Spets. Astrofiz. Obs.,* **7**, 1, 1972.
28. B. E. Markarian, V. A. Lipovetskii, J. A. Stepanian et al., *Soobshch. Spets. Astrofiz. Obs.,* **62**, 5, 1989.
29. B. Takase and N. Miyauchi-Isobe, *Ann. Tokyo Astron. Obs.,* **20**, 237, 1985.
30. J. J. Condon, W. D. Cotton, E. W. Greisen et al., *Astron. J.,* **115**, 1693, 1998.
31. H. Arp, *Astrophys. J. Suppl. Ser.,* **46**, 75, 1981.
32. M. A. Malkan and L. K. Hunt, in *"Coevolution of Black Holes and Galaxies",* Carnegie Observatories Astrophysics Series, Edited by L. C. Ho, Pasadena, 2004.
33. T. H. Jarrett, T. Chester, R. Cutri et al., *Astron. J.,* **119**, 2498, 2000.
34. R. G. Kron, *Astrophys. J. Suppl. Ser.,* **43**, 305, 1980.
35. E. Bertin and S. Arnouts, *Astron. Astrophys. Suppl.,* **117**, 393, 1996.
36. J. M. Carpenter, *Astron. J.,* **121**, 2851, 2001.
37. A. Cabrera-Lavers and F. Garzón, *Astron. Astrophys.,* **403**, 383, 2003.
38. G. Paturel, C. Petit, J. Rousseau, and I. Vauglin, *Astron. Astrophys.,* **405**, 1, 2003.
39. D. S. Mathewson, V. L. Ford, and M. Buchhorn, *Astrophys. J.,* **389**, L5, 1992.
40. H. H. Hurman, *"Modern Factor Analysis",* Univ. of Chicago Press, Chicago, 1967.
41. H. F. Kaiser, *Psychometrika,* **23**, 187, 1958.
42. P. Fouqué, L. Chevallier, M. Cohen et al., *Astron. Astrophys. Suppl.,* **141**, 313, 2000.

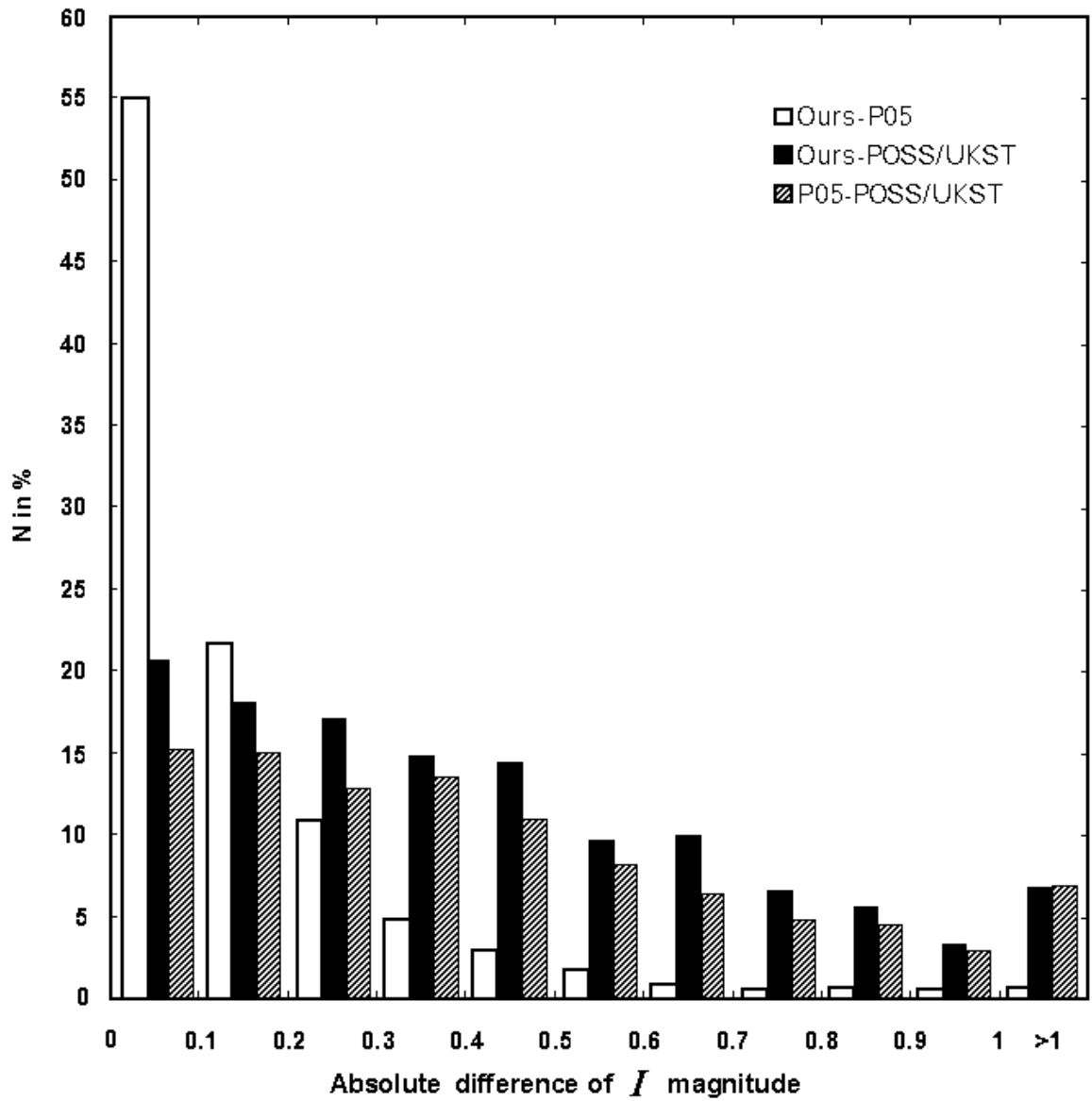

Fig. 1. The percentage distributions of the absolute differences between DENIS and POSS/UKST *I*-band magnitudes.

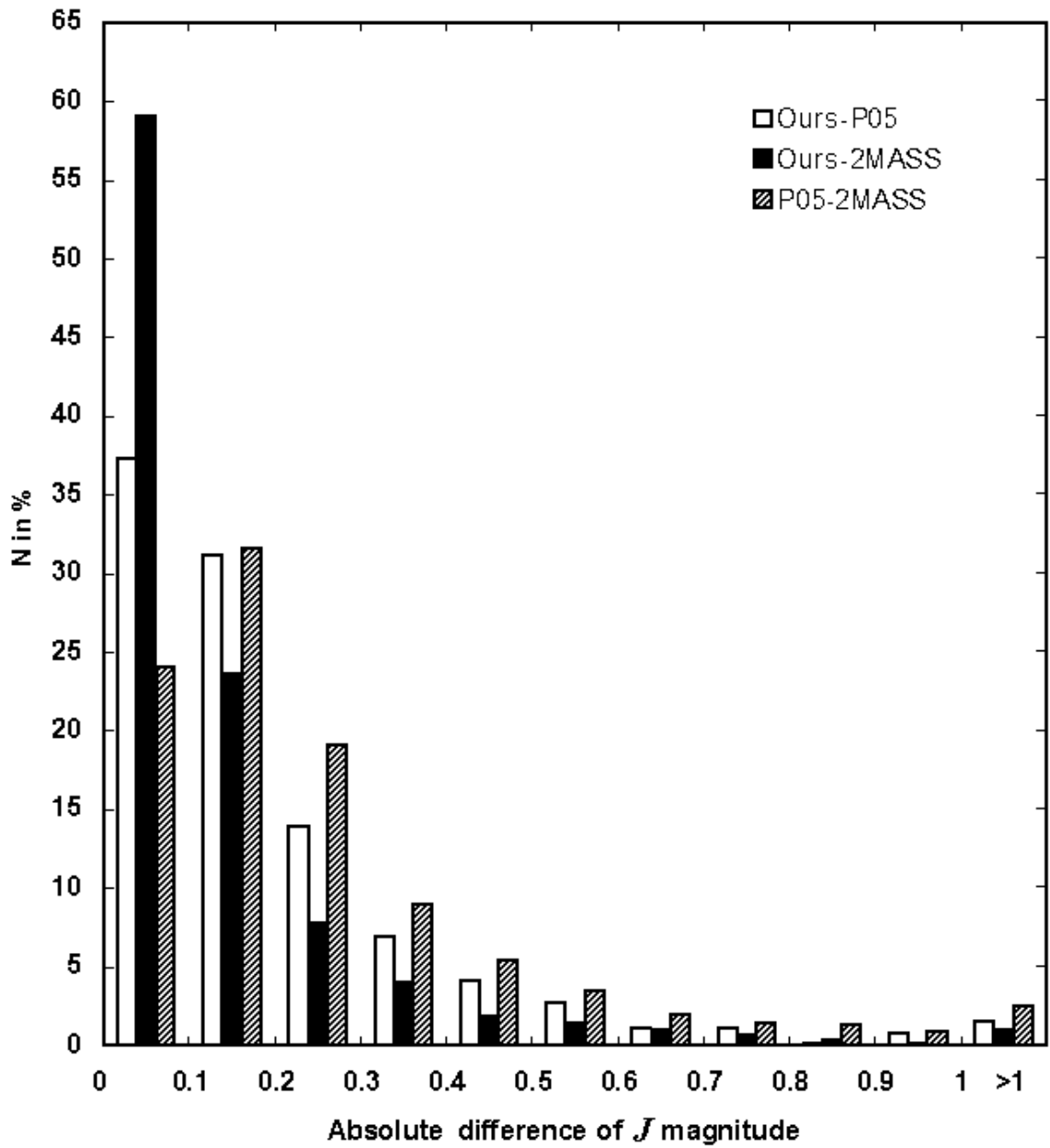

Fig. 2. The percentage distributions of the absolute differences between DENIS and 2MASS *J*-band magnitudes.

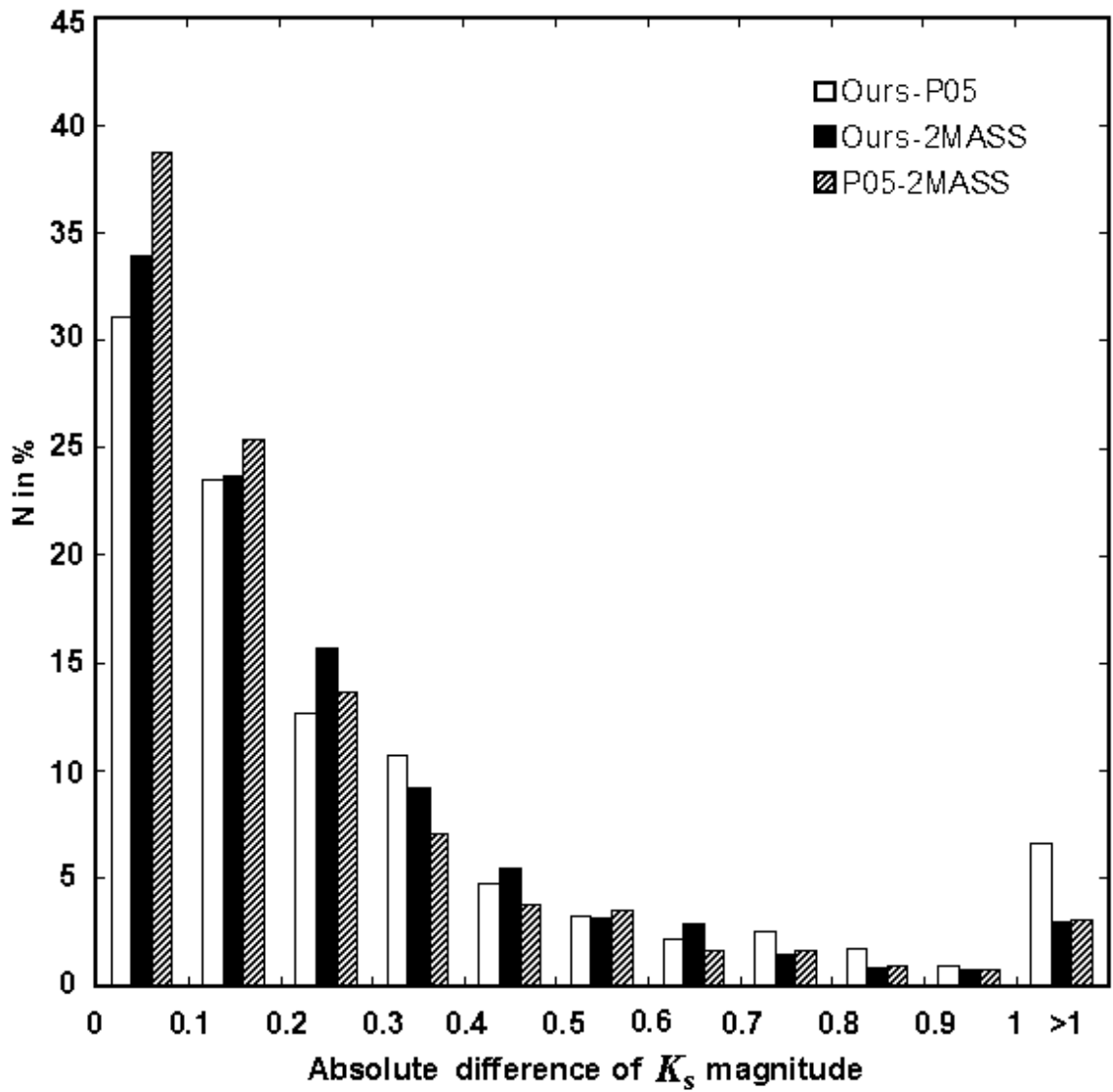

Fig. 3. The percentage distributions of the absolute differences between DENIS and 2MASS $K_s$-band magnitudes.